\documentclass[reprint,twocolumn,superscriptaddress,secnumarabic,amssymb, longbibliography,nobibnotes, aps,pra]{revtex4-1}
\usepackage[T1]{fontenc}
\usepackage[utf8]{inputenc}
\usepackage{amssymb}
\usepackage{amsmath}
\usepackage{color}
\usepackage{xcolor}
\usepackage{hyperref}
\usepackage{verbatim}
\usepackage{graphicx}
\usepackage{gensymb}
\usepackage{comment}

\begin{document}

\preprint{APS/123-QED}

\title{Valency, charge-transfer, and orbital-dependent correlation in bilayer nickelates Nd$_3$Ni$_2$O$_7$}

\author{Daisuke Takegami}
\affiliation{Department of Applied Physics, Waseda University, 3-4-1 Okubo, Shinjuku-ku, Tokyo 169-8555, Japan}
\affiliation{Max Planck Institute for Chemical Physics of Solids, N{\"o}thnitzer Str. 40, 01187 Dresden, Germany}

\author{Takaki Okauchi}
\affiliation{Department of Physics and Electronics, Osaka Metropolitan University, 1-1 Gakuen-cho, Nakaku, Sakai, Osaka 599-8531, Japan}

\author{Edgar Abarca Morales}
\affiliation{Max Planck Institute for Chemical Physics of Solids, N{\"o}thnitzer Str. 40, 01187 Dresden, Germany}

\author{Kouto Fujinuma}
\affiliation{Department of Applied Physics, Waseda University, 3-4-1 Okubo, Shinjuku-ku, Tokyo 169-8555, Japan}

\author{Mizuki Furo}
\affiliation{Department of Physics and Electronics, Osaka Metropolitan University, 1-1 Gakuen-cho, Nakaku, Sakai, Osaka 599-8531, Japan}

\author{Masato Yoshimura}  
\affiliation{National Synchrotron Radiation Research Center, 101 Hsin-Ann Road, 30076 Hsinchu, Taiwan}
\author{Ku-Ding Tsuei}  
\affiliation{National Synchrotron Radiation Research Center, 101 Hsin-Ann Road, 30076 Hsinchu, Taiwan}

\author{Grace A. Pan}
\affiliation{Department of Physics, Harvard University, Cambridge, 02138 USA}

\author{Dan Ferenc Segedin}
\affiliation{Department of Physics, Harvard University, Cambridge, 02138 USA}

\author{Qi Song}
\affiliation{Department of Physics, Harvard University, Cambridge, 02138 USA}

\author{Hanjong Paik}
\affiliation{Platform for the Accelerated Realization, Analysis and Discovery of Interface Materials (PARADIM), Cornell University, Ithaca, 14853 USA}
\affiliation{School of Electrical and Computer Engineering, University of Oklahoma, Norman, OK 73019, USA}
\affiliation{Center for Quantum Research and Technology, University of Oklahoma, Norman, OK 73019, USA}

\author{Charles M. Brooks}
\affiliation{Department of Physics, Harvard University, Cambridge, 02138 USA}

\author{Julia A. Mundy}
\affiliation{Department of Physics, Harvard University, Cambridge, 02138 USA}
\affiliation{John A. Paulson School of Engineering and Applied Sciences, Harvard University, Cambridge, 02138 USA}

\author{Takashi Mizokawa}
\affiliation{Department of Applied Physics, Waseda University, 3-4-1 Okubo, Shinjuku-ku, Tokyo 169-8555, Japan}

\author{Liu Hao Tjeng}
\affiliation{Max Planck Institute for Chemical Physics of Solids, N{\"o}thnitzer Str. 40, 01187 Dresden, Germany}

\author{Berit H. Goodge}
\altaffiliation{corresponding author: Berit.Goodge@cpfs.mpg.de}
\affiliation{Max Planck Institute for Chemical Physics of Solids, N{\"o}thnitzer Str. 40, 01187 Dresden, Germany}

\author{Atsushi Hariki}
\altaffiliation{corresponding author: hariki@omu.ac.jp}
\affiliation{Department of Physics and Electronics, Osaka Metropolitan University, 1-1 Gakuen-cho, Nakaku, Sakai, Osaka 599-8531, Japan}

\date{\today}

\begin{abstract}
We examine the bulk electronic structure of Nd$_3$Ni$_2$O$_7$ using Ni 2$p$ core-level hard x-ray photoemission spectroscopy combined with density functional theory + dynamical mean-field theory. Our results reveal a large deviation of the Ni $3d$ occupation from the formal Ni$^{2.5+}$ valency, highlighting the importance of the charge-transfer from oxygen ligands. We find that the dominant $d^8$ configuration is accompanied by nearly equal contributions from $d^7$ and $d^9$ states, exhibiting an unusual valence state among Ni-based oxides. Finally, we discuss the Ni $d_{x^2-y^2}$ and $d_{z^2}$ orbital-dependent hybridization, correlation and local spin dynamics.
\end{abstract}

\maketitle


Nickel-based complex oxides -- nickelates -- have drawn considerable attention due to their rich phase diagrams of strongly correlated behavior including metal-to-insulator transitions \cite{Mercy17,pardo2011metal,jaramillo2014origins,Bisogni2016,Johnston2014}, density waves \cite{tam2022charge,zhang2020intertwined,chen2024evidence}, magnetism \cite{fowlie2022intrinsic,lu2021magnetic,ortiz2022magnetic}, and superconductivity \cite{Li2019, Pan2022, Sun2023, zhu2024superconductivity, ko2024signatures}.
The Ruddlesden-Popper nickelates R$_{n+1}$Ni$_{n}$O$_{3n+1}$ in particular also provide a platform for exploring the interplay between structural distortions, electronic correlations, and charge-transfer physics \cite{cui2024strain,sun2021electronic}.
{The recent discoveries of superconductivity in the bi- and trilayer compounds ($n = 2, 3$) \cite{Sun2023, zhu2024superconductivity, ko2024signatures, zhou2024ambient} have reinvigorated efforts to understand their electronic structure while introducing a new fundamental challenge: the valency of Ni.}

In square-planar nickelates, such as protypical NdNiO$_2$, the low formal valency of Ni$^{1+}$ accommodates self-doping of holes from the rare-earth 5$d$ states to the Ni ions, introducing additional complexities in low-energy excitations that are absent in high-$T_c$ cuprates superconductors \cite{wang2024experimental,Sakakibara2020,Kitatani2020,Hirayama2020}. 
The bilayer Ruddlesden-Popper R$_3$Ni$_2$O$_7$, by comparison, has a formal valency of Ni$^{2.5+}$. 
The higher Ni valency, i.e., lower Ni 3$d$ levels, avoids self-doping from the rare-earth while a stronger charge-transfer from the O 2$p$ bands may be present~\cite{Dong2024,Chen2024RIXS}, as is often observed in RNiO$_3$ with a formal Ni$^{3+}$ valency~\cite{catalano2018rare,Bisogni2016,Green2016}.
Furthermore, the non-integer Ni$^{2.5+}$ valency poses a fundamental question in modeling the electronic structure, particularly regarding the appropriate starting point for the Ni valency and whether it aligns more closely with Ni$^{2+}$ ($d^8$) or Ni$^{3+}$ ($d^7$). 
It is worth stressing that these two electronic configurations exhibit different atomic multiplet structures and effective hybridization with ligands when forming covalent bonds. 
Thus, this detail represents a fundamental issue that underpins various open questions, including for example the importance of multiorbital physics, the origin of the strong orbital dependence in the mass renormalization, and the absence of a static charge order or disproportionation in the quest to understand superconductivity in the bilayer compounds.

One of the most well-established tools for investigating the electronic states of such complex transition metal systems is core-level photoemission spectroscopy (PES)~\cite{HuefnerBook,DeGroot2008}.  The Ni 2$p$ core-level PES measures the dynamical charge response of low-lying valence electrons to the sudden creation of a highly localized core hole at the excited Ni ion, leading to distinct peaks in the spectrum caused by the charge-transfer (CT) from O 2$p$ and Ni metallic electrons. These are traditionally referred to as local and nonlocal screening, respectively. By analyzing these peaks,
we determine the CT energy parameter that governs valency and effective hybridization with oxygens in CT-type Ni oxides~\cite{Higashi21,Hariki17}. PES is particularly suited for such studies due to its high sensitivity to CT effects, unlike charge-neutral methods such as x-ray absorption spectroscopy~\cite{DeGroot2008}. 
This approach moves beyond limitations of formal electron counting to extract critical information on Ni 3$d$ configuration, correlation effects, and orbital hybridization.

Here, we address this key question by performing core-level hard x-ray photoemission spectroscopy (HAXPES) experiments combined with local density approximation (LDA) + dynamical mean-field thery (DMFT) simulations. 
By making use of the higher probing depth provided by HAXPES, we are able to study the bulk electronic structure of the bilayer nickelates. 
{Measuring the intrinsic Ni 2$p$ core-level spectrum in La$_3$Ni$_2$O$_7$, however, is impossible due to severe overlap between the Ni 2$p_{3/2}$ and La 3$d$ core levels~\cite{Mickevicius2006,Yamagami2021,Takegami2024_LNO}, as shown in Fig.~S3 of Supplementary Material (SM)~\cite{sm}. This overlap has led to conflicting interpretations of the Ni charge state in previous experimental studies~\cite{Liu22,Takegami2024_LNO}, preventing the extraction of a reliable spectrum for theoretical modeling of the R$_3$Ni$_2$O$_7$ system. 
To overcome this issue, we synthesized Nd$_3$Ni$_2$O$_7$ thin films \cite{pan2022synthesis,ferenc2023limits} which enable direct access to the intrinsic Ni 2$p$ spectrum as input for theoretical modeling using the LDA+DMFT method.}
We show that the system presents a dominant $d^8$ configuration, with nearly equal contributions from $d^7$ and $d^9$, differing from typical Ni$^{3+}$ and Ni$^{2+}$ oxides and requiring all configurations to be considered in theoretical models. 
We furthermore explore the distinct hybridization and correlation behaviors in the Ni $d_{x^2-y^2}$ and $d_{z^2}$ orbitals, highlighting the orbital-dependent nature of the electronic structure.

A $\sim30$ nm thin film of Nd$_3$Ni$_2$O$_7$ was grown by ozone-assisted molecular-beam epitaxy (MBE) on a stabilizing LaAlO$_3$ substrate, similarly to previous descriptions \cite{pan2022synthesis, ferenc2023limits}~\footnote{The film was synthesized at a substrate temperature of $670^\circ$~C as measured by a thermocouple and under a chamber pressure of $3\times10^{-6}$~Torr distilled ozone. 
The nickel and neodymium effusion cell temperatures were adjusted to yield a flux of approximately $2 \times 10^{13}$~atoms/cm$^2$s as measured by a quartz crystal microbalance and subsequently fine-tuned using the NdNiO$_3$ calibration procedure detailed in Ref.~\cite{pan2022synthesis}.}. 
{Comparison of the nominal in-plane bulk and substrate lattice constants indicates a resulting compressive strain of $\varepsilon = (a_{bulk} - a_{substrate}) / a_{substrate} \approx - 0.9\%$ \cite{pan2022synthesis, ferenc2023limits}, smaller than that reported in thin films of La$_3$Ni$_2$O$_7$ which exhibit superconductivity \cite{ko2024signatures, zhou2024ambient}.}
{Annular dark-field scanning transmission electron microscopy (ADF-STEM) investigation of the film structure shows good adherence to the bilayer Ruddlesden-Popper structure \cite{sm}.}
HAXPES measurements were performed at the Max-Planck-NSRRC HAXPES end station with MB Scientific A-1 HE analyzer, Taiwan undulator beamline BL12XU of SPring-8~\cite{takegami2019}. 
Photon energies of $h\nu=6.5$~keV and 10~keV with resolutions of around 270~meV and 320~meV respectively were used. 
Soft x-ray photoelectron spectroscopy experiments were performed at the NSRRC-MPI TPS 45A submicron soft x-ray spectroscopy beamline at the Taiwan Photon Source in Taiwan~\cite{HuangMing2019}. The photon energy was set to 1.2~keV, with a resolution of around 150~meV. All measurements were performed at 80~K.

LDA+DMFT calculations were performed with the implementation in Refs.~\cite{Hariki17,Hariki20,Higashi21} for a lattice model spanning Ni 3$d$ and O 2$p$ bands derived from the LDA calculations~\cite{wien2k,wien2wannier,wannier90}. Based on previous DMFT studies for La$_3$Ni$_2$O$_7$~\cite{Shilenko2023,Wan2021,Leonov2020}, we used Hubbard $U$ and Hund's $J$ values of (6.0~eV, 0.95~eV) within the Ni 3$d$ shell. 
{These values give the configuration-averaged Coulomb interaction of $U_{dd}=U-4/9J=5.57$~eV~\cite{Hariki17}.} 
After obtaining a converged DMFT solution with the continuous-time quantum Monte Carlo solver for the Anderson impurity model (AIM), we calculated valence-band spectra and hybridization densities $\Delta(\varepsilon)$ on the real frequency axis, followed by analytical continuation of the self-energy using the maximum entropy method~\cite{jarrell96}. Finally, we computed the Ni 2$p$ core-level PES spectrum from the AIM, incorporating $\Delta(\varepsilon)$ and 2$p$ core orbitals~\cite{Hariki17,Higashi21}. Computational details and the robustness of our results with the model parameters are provided in the SM~\cite{sm}.

In LDA+$X$ methods, a double-counting correction $\mu_{\rm dc}$ needs to be introduced to account for $dd$ interaction effects present in the LDA results. Though a universally accepted form for $\mu_{\rm dc}$ is unavailable~\cite{Kotliar06,Haule15,Karolak10}, it controls Ni 3$d$ levels relative to O 2$p$ bands, thus the CT energy. Following Ref.~\cite{Higashi21}, we use the linear function $\Delta_{dp} = (\varepsilon_{d} - \mu_{\rm dc}) + 7.5 \times U_{\rm dd} - \varepsilon_{p}$, mimicking the CT energy in a cluster model analysis. 
Here, $\varepsilon_{d}$ and $\varepsilon_{p}$ refer to the LDA orbital energies of Ni 3$d$ and O 2$p$, and $U_{dd}$ is the averaged $d$--$d$ interaction. Realistic $\Delta_{dp}$ values are obtained by comparing simulated Ni 2$p$ core-level results to the experimental HAXPES spectrum.

\begin{figure}[]
\begin{center}
\includegraphics[width=0.98\columnwidth]{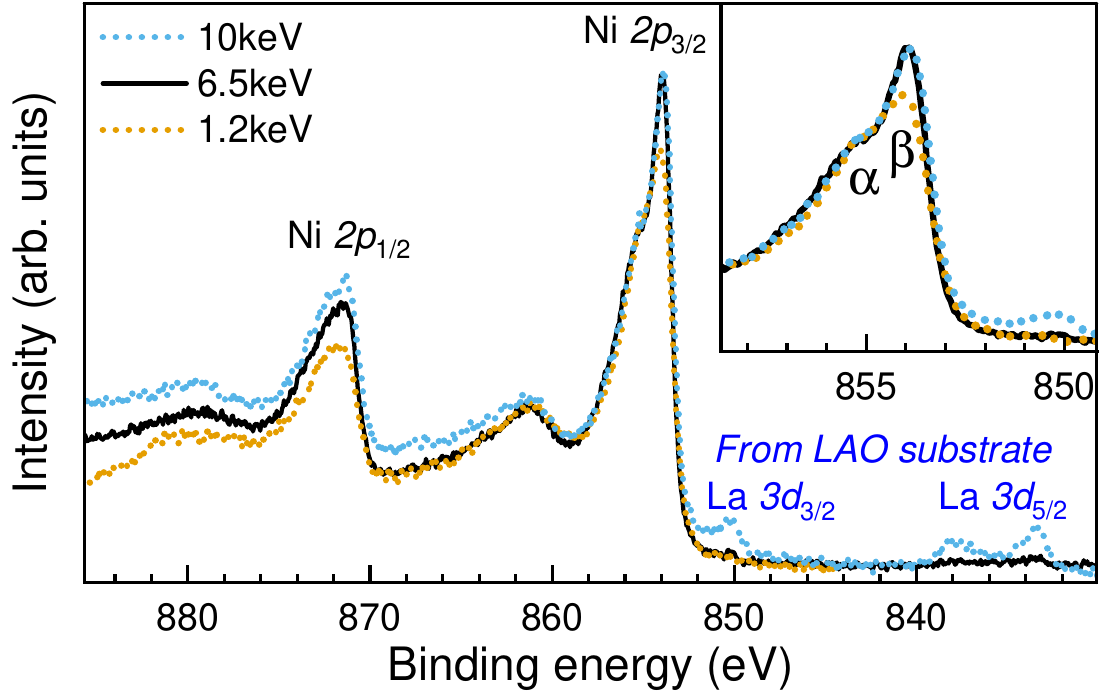}
\end{center}
\caption{
Ni~$2p$ core level spectra measured with photon energies of 10~keV (blue dotted line), 6.5~keV (black line), and 1.2~keV (orange dotted line).
}
\label{Fig_Exp2p}
\end{figure}


Figure~\ref{Fig_Exp2p} shows the Ni~$2p$ core-level photoemission spectra obtained using 10, 6.5, and 1.2~keV photons. 
Overall, the Ni~$2p$ core level displays a set of structures commonly seen in other nickelates~\cite{Hariki2013,Takegami2024_LNO,Yamaguchi2024}. 
The main Ni~$2p_{3/2}$ peak shows two distinct features corresponding to the local ($\alpha$) and nonlocal metallic ($\beta$) screening processes, and a CT satellite at around 861.5~eV. 
Around 17~eV above the Ni~$2p_{3/2}$, we observe a similar structure for the  Ni~$2p_{1/2}$, albeit broader due to the shorter core-hole lifetime. 
Depending on the photon energy, we observe some differences in the spectra. 
In the 10~keV data, there is a weak double peak at 833 and 838~eV, as well as one peak at 850~eV next to the Ni~$2p_{3/2}$, matching the double peak structures of La~$3d$ commonly seen in La-containing perovskite oxides~\cite{Lam1980}, with the La~$3d_{3/2}$ partially overlapping with the Ni~2$p_{3/2}$~\cite{Takegami2024_LNO}. 
These peaks are contributions from the LaAlO$_3$ substrate below the Nd$_3$Ni$_2$O$_7$, which appear when using high photon energies with larger probing depth~\footnote{The probing depth in photoemission experiments depends on the inelastic mean free path of the outgoing photoelectrons, which scales with kinetic energy ($E_K\approx(h\nu-850)$~eV for Ni~$2p$ photoelectrons), and are roughly estimated to be on the order of 15, 10, and 1~nm~\cite{Tanuma2011,HAXPESbook2016} for 10, 6.5, and 1.2~keV photons respectively.}. 
Indeed, at 6.5~keV, the La~$3d$ features are reduced almost completely despite the relative increase of the La~$3d$ cross sections compared to those of Ni~$2p$~\cite{trzhaskovskaya18}, indicating that the probing depth using 6.5~keV is barely enough to reach the substrate. 
This thus confirms the bulk-sensitivity of the measurements and that the collected Ni~2$p$ spectra are well representative of the whole Nd$_3$Ni$_2$O$_7$ film depth, as evidenced by good agreement in the Ni~2$p$ derived features for 10~keV and 6.5~keV despite their different probing depths. 

The surface-sensitive 1.2~keV spectrum, on the other hand, shows noticeable differences compared to the higher probing energy data. 
The electronic structure at the vicinity of the surface is thus clearly distinct from that of the bulk, with the significant reduction of the peak $\beta$ indicating a suppression of the non-local metallic screening at the surface layers. 
Differences between the bulk and surface electronic structure are often known to occur in strongly correlated transition metal oxides~\cite{Taguchi2005,Suga2010,Panaccione2012,Veenstra2013,HAXPESbook2016,Takegami23_LCO,Takegami2024_SFO}, with phenomena like polar surfaces, relaxation or reconstructions in the surface, etc., leading to significantly different properties and band structure near the surface, thus making bulk-sensitivity crucial to ensure that the experimental results are intrinsic and representative of the bulk of the material.
It is important to note the observation of these differences at 1.2~keV, i.e. Ni~$2p$ core level photoelectrons with 300-350~eV kinetic energy, which are expected to have higher probing depths than common VUV valence band ARPES experiments.
\begin{figure}[]
\begin{center}
\includegraphics[width=0.95\columnwidth]{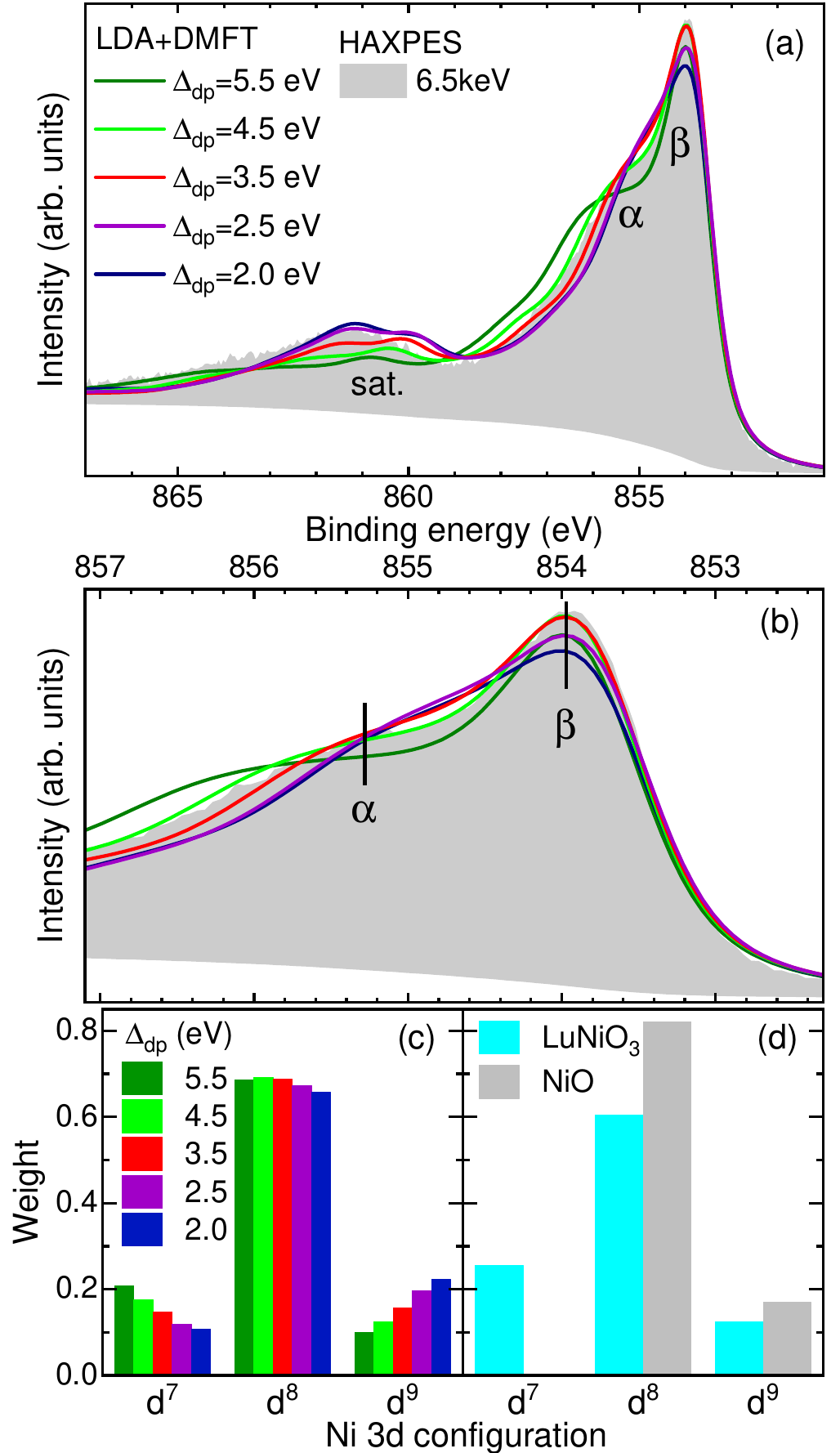}
\end{center}
\caption{(a) LDA+DMFT fit together with the experimental Ni 2$p_{3/2}$ HAXPES spectrum.
(b) Close-up of the Ni 2$p_{3/2}$ $\alpha$ and $\beta$ features.
(c) Atomic configuration histogram of the Ni 3$d$ states in Nd$_3$Ni$_2$O$_7$, computed with the selected $\Delta_{dp}$ values. 
(d) Atomic configuration histograms for reference Ni oxides: metallic LuNiO$_3$ (cyan)~\cite{Winder20} and NiO (gray)~\cite{sm}.}
\label{Fig_2pCalc}
\end{figure}

Having established that the 6.5~eV HAXPES measurements best represent the bulk Nd$_3$Ni$_2$O$_7$ thin film, we use these measurements to determine the $\Delta_{dp}$ parameter by comparison to LDA+DMFT calculations. 
In Fig.~\ref{Fig_2pCalc}(a,b), the Ni~$2p_{3/2}$ spectra are computed for selected $\Delta_{dp}$ values.
{With increasing $\Delta_{dp}$, the ligand levels shift deeper relative to the Fermi level, leading to a larger energy splitting between the local screening $\alpha$ (mainly from nearest-neighboring oxygens) and the nonlocal metallic screening $\beta$ features. In the close-up shown in Fig.~\ref{Fig_2pCalc}(b), we observe that the experimental splitting and $\alpha-\beta$ ratio is best reproduced by $\Delta_{dp} = 3.5$~eV, with $4.5$~eV also yielding a reasonable agreement.
Next, we observe in Fig.~\ref{Fig_2pCalc}(a) that the weight ratio between the main peak and the satellite is also highly sensitive to the $\Delta_{dp}$ value, with the satellite spectral weight decreasing with increasing $\Delta_{dp}$. Here, the best agreement is obtained between $\Delta_{dp} = 3.5$~eV and $2.5$~eV.
These two observations allow us to constrain its realistic value of around $\Delta_{dp} = 3.5$~eV.} 
{In SM~\cite{sm}, we computed the Ni 2$p$ spectrum for not only for different values of $\Delta_{dp}$ but also for different values of $U$. We found that the experimental spectrum is best reproduced by the chosen $U$, although the sensitivity to the precise value of $U$ is not very large.
}

The $\Delta_{dp}$, which measures energy splitting of the Ni 3$d$ and O 2$p$ levels, is a key parameter for the $d$-electron charge states in CT-type systems according to the Zaanen-Sawatzky-Allen (ZSA) diagram~\cite{Zaanen85}. 
In Fig.~\ref{Fig_2pCalc}(c), the Ni~$3d$ charge state is quantified by computing an atomic histogram at the Ni site in DMFT solutions for various $\Delta_{dp}$ values. 
Regardless of $\Delta_{dp}$, the $d^8$ configuration exhibits a dominant peak in the histogram, with a large distribution towards the $d^7$ and $d^9$ configurations which depend on the $\Delta_{dp}$ value: a smaller $\Delta_{dp}$ increases the $d^{9}$ weight, and vice versa. 
At the optimal value determined above, the $d^7$ and $d^{9}$ weights are nearly identical. 
{As shown in Fig.~S6 of SM~\cite{sm}, the $d$-configuration distribution in the atomic histogram is predominantly determined by $\Delta_{dp}$, with negligible influence from the Coulomb interaction $U$, as expected for a system in the CT-type regime of the ZSA diagram.}

\begin{figure}[t]
\begin{center}
\includegraphics[width=0.98\columnwidth]{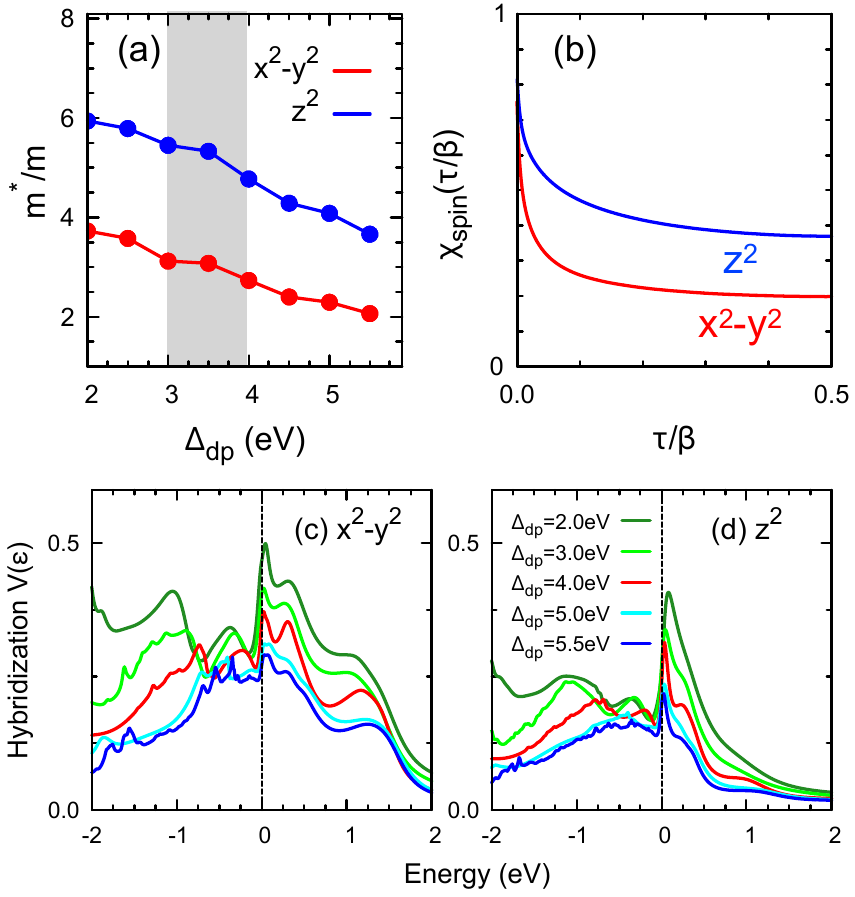}
\end{center}
\caption{(a) The mass enhancement $m^*/m$ of the Ni $d_{x^2-y^2}$ and $d_{z^2}$ states as a function of the $\Delta_{dp}$ parameters. The region in gray shadow indicates the realistic values estimated by the Ni 2$p$ XPS analysis in Fig.~\ref{Fig_2pCalc}. (b) The orbital-diagonal component of the local spin correlation function $\chi_{\rm spin}(\tau)$ with $\Delta_{dp}=3.5$~eV, where $\tau$ represents the imaginary time. The Ni hybridization densities $V_{\gamma}(\omega)$ for (c) the $d_{x^2-y^2}$ state and (d) the $d_{z^2}$ state, calculated with the different $\Delta_{dp}$ values. All results in the panels are calculated using the LDA+DMFT method at $T=300$~K.}
\label{Fig_mass}
\end{figure}

The charge state contrasts with that of divalent or trivalent Ni oxides. 
In Fig.~\ref{Fig_2pCalc}(d), the histograms for NiO and metallic LuNiO$_3$, which are prototype systems of formally Ni$^{2+}$ and Ni$^{3+}$ oxides, respectively, are shown. 
The reference data are taken from Ref.~\cite{Winder20,foot_lunio3} and an additional DMFT simulation in SM~\cite{sm}. 
NiO exhibits a predominant $d^8$ peak with a distribution toward the $d^9$ state. 
In the high-valency LuNiO$_3$, with deeper Ni 3$d$ levels, the CT energy is, as in other formally Ni$^{3+}$ oxides~\cite{Mizokawa1995,Bisogni2016,Green2024}, small or even negative, facilitating CT from O 2$p$ states and resulting in a dominant $d^8$ state. 
The larger weight of the $d^7$ state compared to that of $d^9$ reflects its high formal valency.
For an optimal $\Delta_{dp}$ of 3.5 eV, the charge state of Nd$_3$Ni$_2$O$_7$, with formally Ni$^{+2.5}$, is qualitatively different from these reference Ni oxides, and all the $d^7$--$d^{9}$ electronic configurations need to be considered when modeling its electronic structure. 
The mean Ni occupation of nearly $d^8$, deviating from the expected $d^{\sim 7.5}$ for a formal Ni$^{+2.5}$ valency, suggests that charge-transfer from O 2$p$ states is significant in Nd$_3$Ni$_2$O$_7$.
Note as $\Delta_{dp}$ is reduced (increased), the $d$ configuration shifts to more closely resemble either divalent (trivalent) cases. 
{The $d$ occupation in LuNiO$_3$~\cite{Winder20} is 7.84, exhibiting a larger deviation from its formal valency than that in Nd$_3$Ni$_2$O$_7$, which is consistent with $\Delta_{dp}$ of 3.5~eV in Nd$_3$Ni$_2$O$_7$ being moderate.}

{Our results are consistent with the recent study \cite{Takegami2024_LNO} which indicated the absence of charge disproportionation (CD) signatures in La$_3$Ni$_2$O$_7$.}
Our LDA+DMFT calculations do not indicate any instability toward a CD~\footnote{We performed additional LDA+DMFT calculations for the experimental structure, allowing for a checkerboard pattern of CD as observed in $R$NiO$_3$.}. 
Its absence would be unsurprising, given that the Ni $d$ charge states (Fig.~\ref{Fig_2pCalc}(b)(c)) as well as the crystal structure of $R$$_3$Ni$_2$O$_7$ differ from those of $R$NiO$_3$, where CD is widely observed.
The standard picture of CD in $R$NiO$_3$ depends on a delicate balance between the three-dimensional tilting pattern of NiO$_6$ octahedra and the correlated Ni 3$d$ charge states~\cite{Park12,Georgescu19,Ruppen15,Peil19,Lee11,Mercy17},
while the bilayer Ruddlesden-Popper compounds should host distinct octahedral distortions due to their reduced structural dimensionality. 
Theoretical models suggest these distortions are likely tied to the emergence of superconductivity under suitable pressure conditions \cite{geisler2024structural} particularly in relation to orbital anisotropy, which we return to below.

The clarification of the Ni $d$ state has implications for understanding low-energy excitations. Although the Ni $d^8$ valence configuration weight remains almost unchanged over a wide range of $\Delta_{dp}$, as shown in Fig.~\ref{Fig_2pCalc}(c), this does not necessarily imply that low-energy excitations are unaffected. In Fig.~\ref{Fig_mass}(a), the quasi-particle mass enhancement $m^*/m$ of the Ni $d_{x^2-y^2}$ and $d_{z^2}$ states was calculated from the converged DMFT self-energies $\Sigma(i\omega_n)$, as $m^*/m=Z^{-1}=[1-\partial {\rm Im}\Sigma(i\omega)/\partial \omega]_{\omega\rightarrow 0^+}$, {where the $m$ is noninteracting band mass and $Z$ is the renormalization factor}. 
In addition to strong orbital dependence, the $m^*/m$ of the two orbitals exhibits a pronounced dependence on the $\Delta_{dp}$ value, as the O 2$p$ states are present near the Fermi energy $E_F$ in this CT-type oxide and thus directly influence the degree of the localization of the Ni 3$d$ electrons, see SM~\cite{sm} for the $\Delta_{dp}$ dependence of the DMFT valence-band spectra. 
{A recent ARPES study on La$_3$Ni$_2$O$_7$ estimated $m^*/m$ to be in the range of 1–3 for the $d_{x^2-y^2}$ state and 5–8 for the $d_{z^2}$ state by rescaling DFT and DFT+$U$ bands to match ARPES results in part of the Brillouin zone~\cite{Yang24}. The calculated $m^*/m$ values for the $\Delta_{dp}$ range that reproduces the Ni 2$p$ data in Fig.~\ref{Fig_2pCalc}, highlighted in gray in Fig.\ref{Fig_mass}(a), are comparable to the reported values.}

Orbital dependence has been at the center of discussions in the bilayer nickelates~\cite{Shilenko23,Sakakibara24,Yang24,Lechermann23,Cao24,Ouyang24}. 
To gain an insight into it, in Figs.~\ref{Fig_mass}(c)(d), we compare the Ni hybridization densities $V_{\gamma}(\omega)$ ($\gamma=x^2-y^2, z^2$), which represent the exchange amplitude of an electron between the local $d_\gamma$ state and lattice. 
The sharp evolution near $E_F$ in $V_{\gamma}(\omega)$ as a function of $\Delta_{dp}$ accounts for the sensitivity of the metallic screening feature $\beta$ in Ni 2$p$ core-level PES to the $\Delta_{dp}$ values in Fig.~\ref{Fig_2pCalc}(a). 
Overall, the Ni $d_{x^2-y^2}$ state exhibits stronger hybridization with the low-energy states compared to the Ni $d_{z^2}$ state, 
which is likely responsible for the weaker $m^*/m$ for the former orbital than the latter one. 
This observation is consistent with the orbital-dependent spin screening encoded in 
the time-dependent local spin correlation function $\chi_{\rm spin}(\tau)$, calculated for the LDA+DMFT result with the optimal $\Delta_{dp}$ in Fig.~\ref{Fig_mass}(b). 
{$\chi_{\rm spin}(\tau)$ characterizes the dynamics of the local spin, influenced by hybridization with the crystal and many-body correlations~\cite{Takegami22,Hariki17_2,Watzenbock20}. We find a finite and nearly identical response at $\tau=0^+$ for both orbitals, indicating that an instantaneous spin of approximately $S=1/2$ is present in each. However, the Ni $d_{x^2-y^2}$ electron undergoes faster spin screening than the $d_{z^2}$ orbital over time, reflecting the orbital-dependent hybridization.}
{In Appendix~A, we demonstrate through additional LDA+DMFT AIM calculations that the orbital-dependent hybridization near $E_F$ is also important for the formation of the $\beta$ feature observed in the experimental Ni 2$p$ core-level spectrum in Fig.~\ref{Fig_2pCalc}(a).}

{The $m^*/m$ values obtained for the ambient pressure phase in this study differ from those calculated for the high-pressure phase in several theoretical studies of La$_3$Ni$_2$O$_7$~\cite{Shilenko23,Cao24}, e.g.~$m^*/m \sim 3$ and 2.3 for the $d_{x^2-y^2}$ and $d_{z^2}$ states, respectively in Ref.~\cite{Shilenko23}. As expected, $m^*/m$ is reduced in the high-pressure phase due to the increased bandwidth of the Ni 3$d$ states compared to the ambient pressure, while the orbital-dependent localization behavior persists in the high pressure phase.}
{In future, extending the approach described here to epitaxially strained thin films could provide further insight to the evolution of such parameters with structural tuning. }



In summary, we have characterized the bulk electronic structure of a bilayer nickelate Nd$_3$Ni$_2$O$_7$ thin film using Ni 2$p$ core-level hard x-ray photoemission spectroscopy (HAXPES) combined with LDA+DMFT simulations. Comparison of the surface-sensitive PES and bulk-sensitive HAXPES measurements show significant differences between the surface and bulk electronic structures in this compound. We experimentally observed both local and nonlocal screening features in the Ni 2$p_{3/2}$ core-level spectrum, which was not feasible in La-based samples studied so far due to overlapping La 3$d$ and Ni 2$p$ core levels. Guided by the observed core-level features, we performed parameter optimization in the LDA+DMFT calculations and determined the charge-transfer energy, a key parameter for the Ni valency and the hybridization with the oxygen ligands in this charge-transfer-type system. Our results show a dominant $d^8$ (Ni$^{2+}$) configuration ($\sim 70\%$), with nearly equal contributions from $d^7$ (Ni$^{3+}$) and $d^9$ (Ni$^{1+}$) ones. This charge distribution differs from typical Ni$^{3+}$ oxides like RNiO$_3$ and Ni$^{2+}$ oxides like NiO, requiring all configurations to be considered in theoretical models. Moreover, the Ni $d_{x^2-y^2}$ orbital exhibited stronger hybridization compared to the $d_{z^2}$ orbital, leading to distinct correlation (mass renormalization) and spin dynamic behaviors in the two orbitals. This highlights the orbital-dependent nature of the electronic structure and correlations in this system.

\section*{Acknowledgements}

We thank M. Kitatani for valuable discussions. D.T.~acknowledges the support by the Deutsche Forschungsgemeinschaft (DFG, German Research Foundation) under the Walter Benjamin Programme, Projektnummer 521584902. 
A.H.~was supported by JSPS KAKENHI Grant Numbers 21K13884, 23K03324, 23H03817. 
B.H.G.~was supported by Schmidt Science Fellows in partnership with the Rhodes Trust. 
We acknowledge the support for the measurements from the Max Planck-POSTECH-Hsinchu Center for Complex Phase Materials. G.A.P. and D.F.S. are primarily supported by U.S. Department of Energy (DOE), Office of Basic Energy Sciences, Division of Materials Sciences and Engineering, under Award No. DE SC0021925; and by NSF Graduate Research Fellowship Grant No. DGE-1745303.  G.A.P. acknowledges additional support from the Paul \& Daisy Soros Fellowship for New Americans.  Q.S. was supported by the Science and Technology Center for Integrated Quantum Materials, NSF Grant No. DMR-1231319.  J.A.M. acknowledges support from the U.S. Department of Energy (DOE), Office of Basic Energy Sciences, Division of Materials Sciences and Engineering, under Award No. DE SC0021925.  Materials growth was supported by PARADIM under National Science Foundation (NSF) Cooperative Agreement No. DMR-2039380. The computation in this work has been done using the facilities of the Supercomputer Center, the Institute for Solid State Physics, the University of Tokyo.

\appendix

\section{Orbital dependent hybridization effect in Ni 2$p$ spectrum}

\begin{figure}[h]
\begin{center}
\includegraphics[width=0.95\columnwidth]{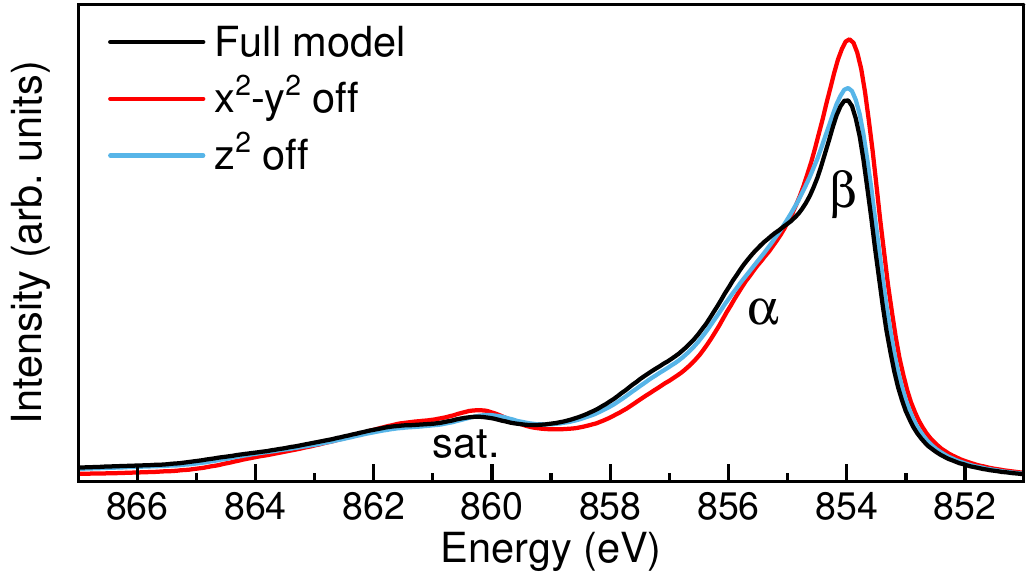}
\end{center}
\caption{Ni 2$p$ core-level PES spectra calculated using the LDA+DMFT AIM method for the full (black) and modified models with no hybridization of the $d_{x^2-y^2}$ orbital (red) and the $d_{z^2}$ orbital (blue) with metallic states near $E_F$. See text for detailed explanations of models.}
\label{Fig_demo}
\end{figure}

{In Fig.~\ref{Fig_demo}, we calculate the PES spectra using the LDA+DMFT AIM with modified $V_\gamma(\omega)$, where the hybridization densities within the [$-1.0$~eV, 1.0~eV] window in Figs.~\ref{Fig_mass}(c)(d) are manually set to zero when simulating the PES final states, meaning that CT screening from the metallic states near $E_F$ does not appear in the simulated spectra. The spectra exhibit intensity modulations of the metallic screening feature $\beta$, with the large suppression of $\beta$ being more closely related to the hybridization of the Ni $d_{x^2-y^2}$ state rather than the $d_{z^2}$ state. This additional piece of evidence derived from the Ni~$2p$ core-level PES spectra further confirms that there is an orbital-dependent hybridization, with the Ni $d_{x^2-y^2}$ state exhibiting a stronger hybridization with the low-energy states compared to the Ni $d_{z^2}$.}

%

\end{document}


\title{Supplementary Material of ``Valency, charge-transfer, and orbital-dependent correlation in bilayer nickelates Nd$_3$Ni$_2$O$_7$''}

\author{Daisuke Takegami}
\affiliation{Department of Applied Physics, Waseda University, 3-4-1 Okubo, Shinjuku-ku, Tokyo 169-8555, Japan}
\affiliation{Max Planck Institute for Chemical Physics of Solids, N{\"o}thnitzer Str. 40, 01187 Dresden, Germany}

\author{Takaki Okauchi}
\affiliation{Department of Physics and Electronics, Osaka Metropolitan University, 1-1 Gakuen-cho, Nakaku, Sakai, Osaka 599-8531, Japan}

\author{Edgar Abarca Morales}
\affiliation{Max Planck Institute for Chemical Physics of Solids, N{\"o}thnitzer Str. 40, 01187 Dresden, Germany}

\author{Kouto Fujinuma}
\affiliation{Department of Applied Physics, Waseda University, 3-4-1 Okubo, Shinjuku-ku, Tokyo 169-8555, Japan}

\author{Mizuki Furo}
\affiliation{Department of Physics and Electronics, Osaka Metropolitan University, 1-1 Gakuen-cho, Nakaku, Sakai, Osaka 599-8531, Japan}

\author{Masato Yoshimura}  
\affiliation{National Synchrotron Radiation Research Center, 101 Hsin-Ann Road, 30076 Hsinchu, Taiwan}
\author{Ku-Ding Tsuei}  
\affiliation{National Synchrotron Radiation Research Center, 101 Hsin-Ann Road, 30076 Hsinchu, Taiwan}

\author{Grace A. Pan}
\affiliation{Department of Physics, Harvard University, Cambridge, 02138 USA}

\author{Dan Ferenc Segedin}
\affiliation{Department of Physics, Harvard University, Cambridge, 02138 USA}

\author{Qi Song}
\affiliation{Department of Physics, Harvard University, Cambridge, 02138 USA}

\author{Hanjong Paik}
\affiliation{Platform for the Accelerated Realization, Analysis and Discovery of Interface Materials (PARADIM), Cornell University, Ithaca, 14853 USA}
\affiliation{School of Electrical and Computer Engineering, University of Oklahoma, Norman, OK 73019, USA}
\affiliation{Center for Quantum Research and Technology, University of Oklahoma, Norman, OK 73019, USA}

\author{Charles M. Brooks}
\affiliation{Department of Physics, Harvard University, Cambridge, 02138 USA}

\author{Julia A. Mundy}
\affiliation{Department of Physics, Harvard University, Cambridge, 02138 USA}
\affiliation{John A. Paulson School of Engineering and Applied Sciences, Harvard University, Cambridge, 02138 USA}

\author{Takashi Mizokawa}
\affiliation{Department of Applied Physics, Waseda University, 3-4-1 Okubo, Shinjuku-ku, Tokyo 169-8555, Japan}

\author{Liu Hao Tjeng}
\affiliation{Max Planck Institute for Chemical Physics of Solids, N{\"o}thnitzer Str. 40, 01187 Dresden, Germany}

\author{Berit H. Goodge}
\altaffiliation{corresponding author: Berit.Goodge@cpfs.mpg.de}
\affiliation{Max Planck Institute for Chemical Physics of Solids, N{\"o}thnitzer Str. 40, 01187 Dresden, Germany}

\author{Atsushi Hariki}
\altaffiliation{corresponding author: hariki@omu.ac.jp}
\affiliation{Department of Physics and Electronics, Osaka Metropolitan University, 1-1 Gakuen-cho, Nakaku, Sakai, Osaka 599-8531, Japan}

\date{\today}

\maketitle

\renewcommand{\thefigure}{S\arabic{figure}}

{
\section{Scanning transmission electron microscopy of the N\lowercase{d}$_3$N\lowercase{i}$_2$O$_7$ thin film}
}
\begin{figure}[h!]
\begin{center}
\includegraphics[width=0.5\columnwidth]{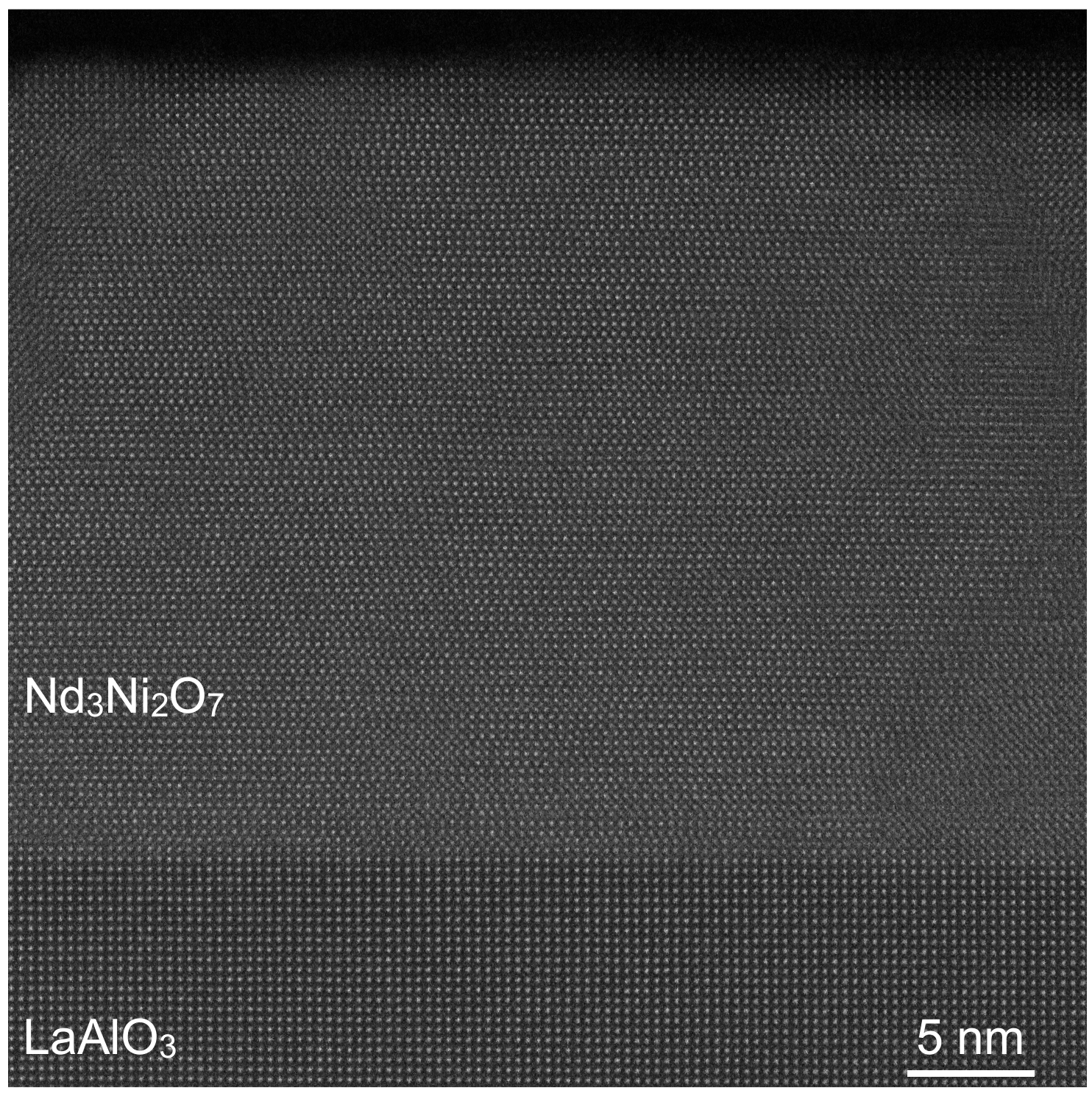}
\end{center}
\caption{
{ Annular dark-field scanning transmission electron microscopy (ADF-STEM) image of the Nd$_3$Ni$_2$O$_7$ thin film measured by HAXPES. The film exhibits overall high crystalline quality with predominant adherence to the bilayer Ruddlesden-Popper stacking structure. 
A cross-sectional lamella was prepared using the standard focused ion beam (FIB) lift-out procedure using a Thermo Fisher Scientific Helios NanoLab G3 CX FIB. STEM images were collected in high-angle annular dark-field (HAADF) configuration on a double aberration-corrected JEOL ARM300F operating at 300 kV with a probe convergence semiangle of 30 mrad. }
}
\label{Fig_STEM}
\end{figure}

{
\section{Additional experimental spectra}
}
\begin{figure}[]
\begin{center}
\includegraphics[width=0.6\columnwidth]{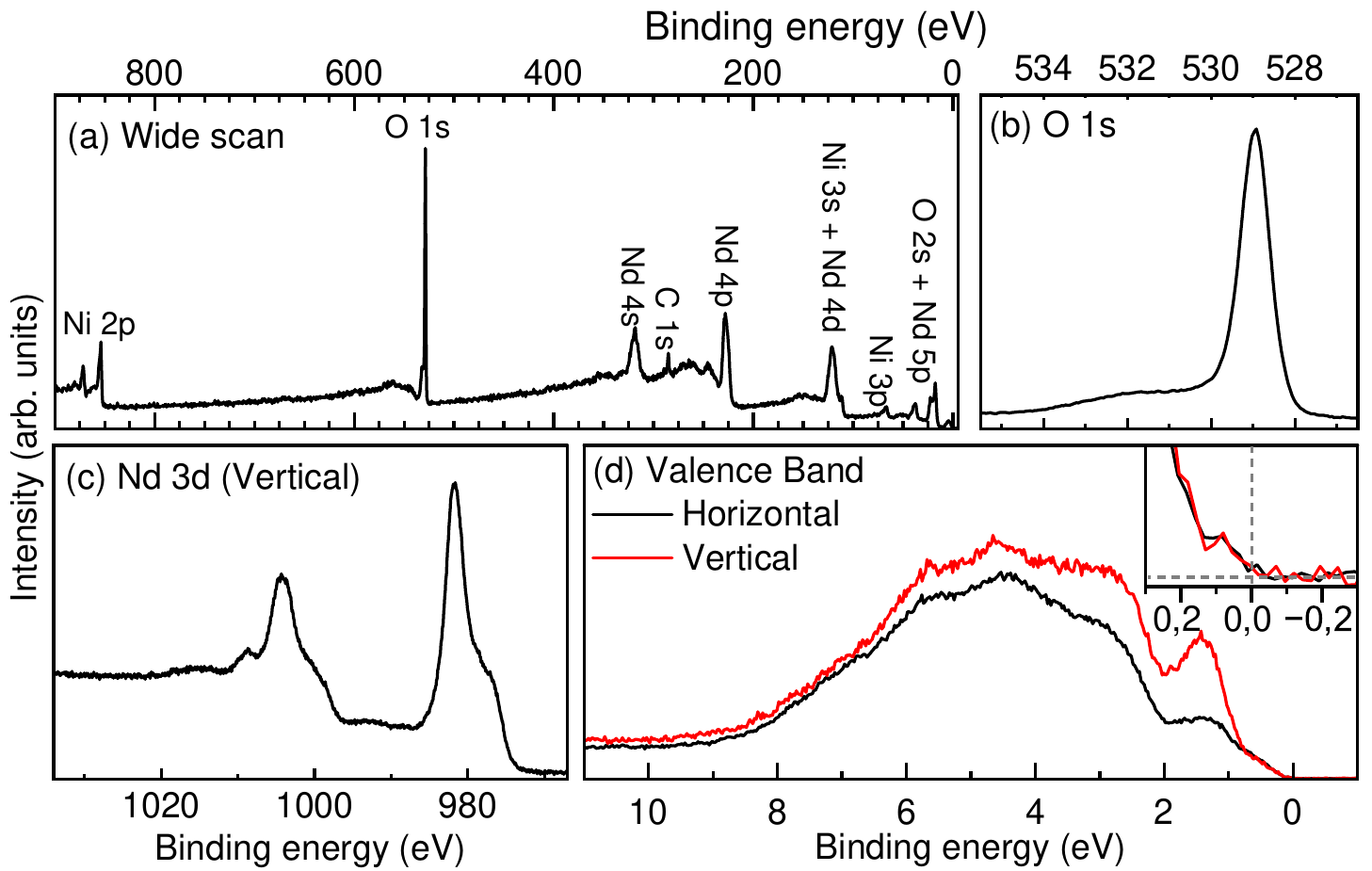}
\end{center}
\caption{
HAXPES wide survey scan (a),  O~$1s$ (b), and Nd~$3d$ (c) core level spectra measured with a photon energy of 6.5~keV. The Nd~$3d$ measurements were performed in the vertical geometry~\cite{takegami2019} in order to suppress the overlapping Ni~$2s$ contributions. {(d) Valence band spectra measured in the horizontal and vertical geometries. The inset shows a close-up near the Fermi energy.}
}
\label{Fig_Others}
\end{figure}
Fig.~\ref{Fig_Others}(a) shows the wide survey scan of Nd$_3$Ni$_2$O$_7$. All observed peaks are consistent with the expected composition of the film, except for the C~$1s$ from the adsorbed impurities on the untreated surface of the film. The O~$1s$, in Fig.~\ref{Fig_Others}(b), displays one single sharp peak from the oxide, with only a weak higher energy tail from the surface impurities, indicating that the obtained HAXPES data is representative of the intrinsic Nd$_3$Ni$_2$O$_7$ contributions. The Nd~$3d$ core level multiplet structure is consistent with that of Nd$_2$O$_3$~\cite{Kotani1992}. {Figure~\ref{Fig_Others}(d) shows the experimental valence band spectra. Overall, the HAXPES spectra is dominated by the O~$2p$ bands through their hybridization with the RE~$5p$, which are highly enhanced due to the large photoionization cross-sections~\cite{takegami2019}. The stronger polarization dependence for the peak at 1.5~eV indicates its dominant Ni~$3d$ character (from the Ni~$3d$ $t_{2g}$, as discussed below), while the weaker dependence closer to the Fermi energy is the result of the stronger mixing of the $e_g$ states with the O~$2p$. At the Fermi energy, a finite but suppressed spectral weight is observed.}\\

{
\section{Comparison to L\lowercase{a}$_3$N\lowercase{i}$_2$O$_7$}
}
\begin{figure}[h]
\begin{center}
\includegraphics[width=0.75\columnwidth]{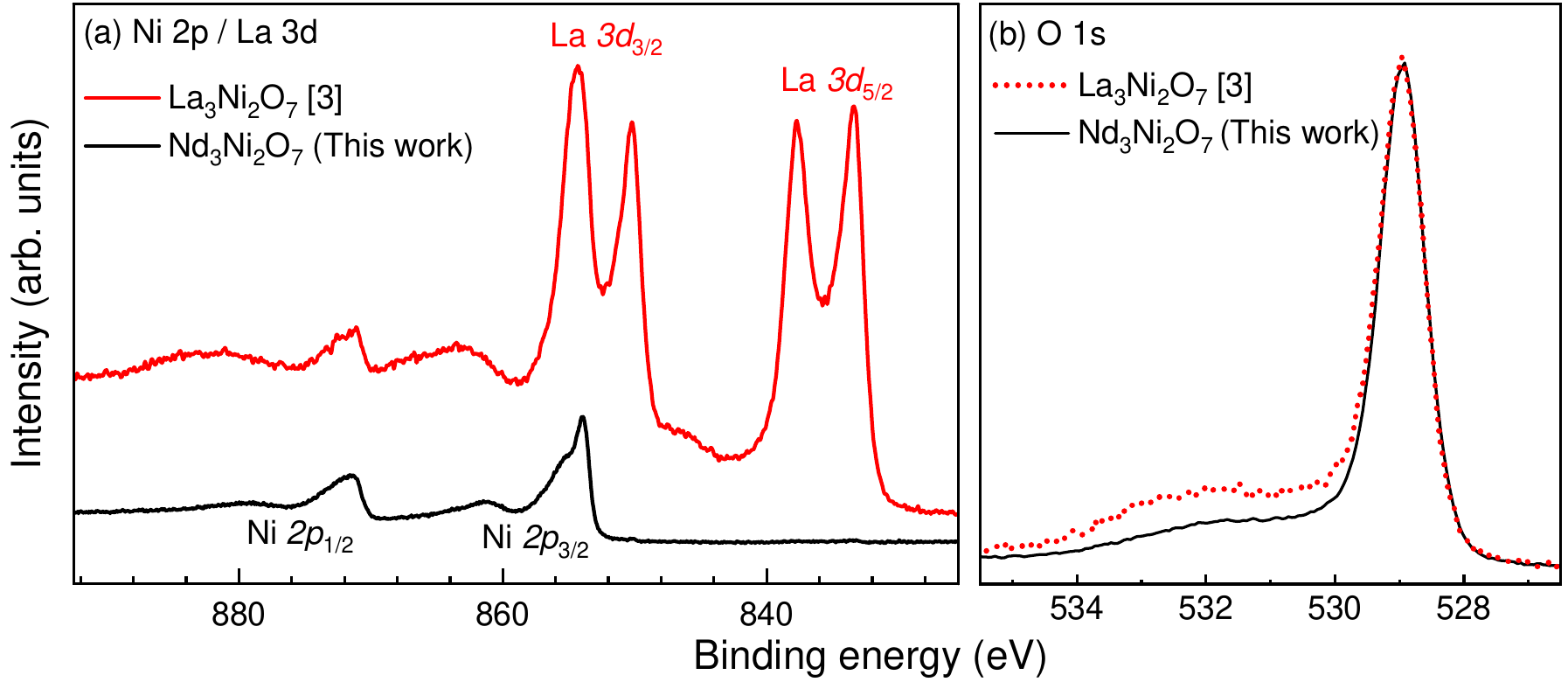}
\end{center}
\caption{{Comparison of the Nd$_3$Ni$_2$O$_7$ experimental HAXPES spectra used in this work with the La$_3$Ni$_2$O$_7$ from Ref.\cite{Takegami2024_LNO} for the Ni~2$p$ and La~3$d$ region (a), and the O~$1s$ (b). In La$_3$Ni$_2$O$_7$, the spectral weight of the La~3$d$ is significantly larger than that of the Ni~2$p$.}}
\label{Fig_LNO}
\end{figure}
{Figure~\ref{Fig_LNO}(a) shows a comparison of the experimental HAXPES spectra in the vicinity of the Ni~2$p$ core level binding energy for Nd$_3$Ni$_2$O$_7$ and La$_3$Ni$_2$O$_7$ (from Ref.~\cite{Takegami2024_LNO}). In the presence of La, two sets of double peaks show up in the region due to the La~3$d_{3/2}$ and La~3$d_{5/2}$ contributions, with the La~3$d_{3/2}$ signal overlapping, and largely dominating the Ni~2$p_{3/2}$ region. As a result, any attempt to estimate the Ni~2$p_{3/2}$ spectral shape in La-containing systems such as La$_3$Ni$_2$O$_7$ requires some form of data treatment method. Such procedures are highly sensitive to the assumptions used for the overlapping La~3$d$ complex spectral lineshape and are thus not as reliable as raw experimental spectra.

Figure~\ref{Fig_LNO}(b) shows a comparison of the O~$1s$ spectra from the Nd$_3$Ni$_2$O$_7$ in this study and the polycrystalline La$_3$Ni$_2$O$_7$ from Ref.~\cite{Takegami2024_LNO}. Oxygen deficiencies are usually manifested in the form of an asymmetric shape of the O~$1s$ main peak~\cite{Takegami2024_SFO}.  In both cases displayed in Fig.~\ref{Fig_LNO}(b), the O~$1s$ main peaks are sharp and Guassian-like, with La$_3$Ni$_2$O$_7$ displaying a slightly more asymmetric shape at around $529-530$~eV. This suggests a similar, or lower amount of oxygen deficiency in the measured Nd$_3$Ni$_2$O$_7$ compared to the polycrystalline La$_3$Ni$_2$O$_7$ from Ref.~\cite{Takegami2024_LNO}, reported to have $\delta\approx 0.07$.
}

\section{LDA+DMFT calculations and model-parameter dependence of spectra}

We calculate the DFT bands using the local density approximation (LDA) for the exchange-correlation potential with the WIEN2K package~\cite{wien2k}. The DFT calculations are performed for La$_3$Ni$_2$O$_7$ (using its crystal structure at ambient pressure~\cite{Voronin01}), a representative compound of the R$_3$Ni$_2$O$_7$ series. 
Since the La/Nd 5$d$ states are well-separated above the Fermi level in the R$_3$Ni$_2$O$_7$ compounds, their contributions to the spectra and physical quantities discussed in the main text can be safely neglected. 
{Moreover, the hybridization between Ni 3$d$ and Nd 4$f$ orbitals is expected to be substantially weaker than that between Ni 3$d$ and O 2$p$. As a result, the relationship between the Ni 2$p$ core-level spectrum and Ni valency in Fig.~2 of the main text remains unaffected by the presence of Nd 4$f$ states.}
The calculated DFT bands are subsequently mapped onto a tight-binding (TB) model explicitly including all Ni 3$d$ and O 2$p$ bands, using the wien2wannier and wannier90 packages~\cite{wannier90,wien2wannier}. The Ni 3$d$ electron self-energy is computed using the auxiliary Anderson impurity model (AIM) in the DMFT self-consistent calculations, employing the continuous-time hybridization expansion Monte Carlo method~\cite{Werner2006a}. Spectral functions and hybridization densities $\Delta(\omega)$ on the real-frequency axis are obtained by analytically continuing the self-energy $\Sigma(\omega)$ via the maximum entropy method~\cite{jarrell96}. The Ni 2$p$ core-level HAXPES spectra are calculated from the AIM, incorporating the DMFT hybridization densities $\Delta(\omega)$, as detailed in Refs.~\cite{Hariki17,Winder20,Ghiasi19}. In this step, the AIM is extended to include the core orbitals and their interactions ($U_{dc}$) with the valence electrons.

Figure~\ref{Fig_VB_dmft} shows the LDA+DMFT valence-band spectral intensities of Nd$_3$Ni$_2$O$_7$ calculated for selected $\Delta_{dp}$ parameters. 
{Besides the Ni $d_{x^2-y^2}$ (red) and $d_{z^2}$ (blue), we find the sizable weights of the O 2$p$ states near the Fermi energy. Note that the spectral intensities of the O 2$p$ states are scaled by a factor of 0.5 in the panels. The bandwidth of the $x^2-y^2$ state is wider than that of the $z^2$ state. The center of the Ni $t_{2g}$ band is approximately 1.5~eV, in good agreement with the experimental VB data mentioned above.}
In Fig.~\ref{Fig_hist_dmft}, we present the Ni atomic configuration histograms computed for different Hubbard $U$ parameters. We observe that the histogram depends strongly on the $\Delta_{dp}$ value, but is rather insensitive to changes in the $U$ parameter. This behavior is consistent with the system being of a charge-transfer type system. To further support the choice of the $U$ parameter in the main text, Fig.~\ref{Fig_XPS_dmft} shows the $U$ dependence of the Ni 2$p$ core-level XPS spectrum for a range of $\Delta_{dp}$ values in the LDA+DMFT AIM result. We find good agreement with the experimental spectrum for both the Ni 2$p_{3/2}$ main line (ML) and the satellite (Sat.) when $U=6.0$~eV is used. In the 2$p$ core-level XPS and XAS simulations, we use a core-hole potential value of $U_{dc}=1.2 \times U_{dd}$, a well-established empirical relation for 3$d$ transition metal oxides~\cite{Hariki17}.

\begin{figure}[]
\begin{center}
\includegraphics[width=0.58\columnwidth]{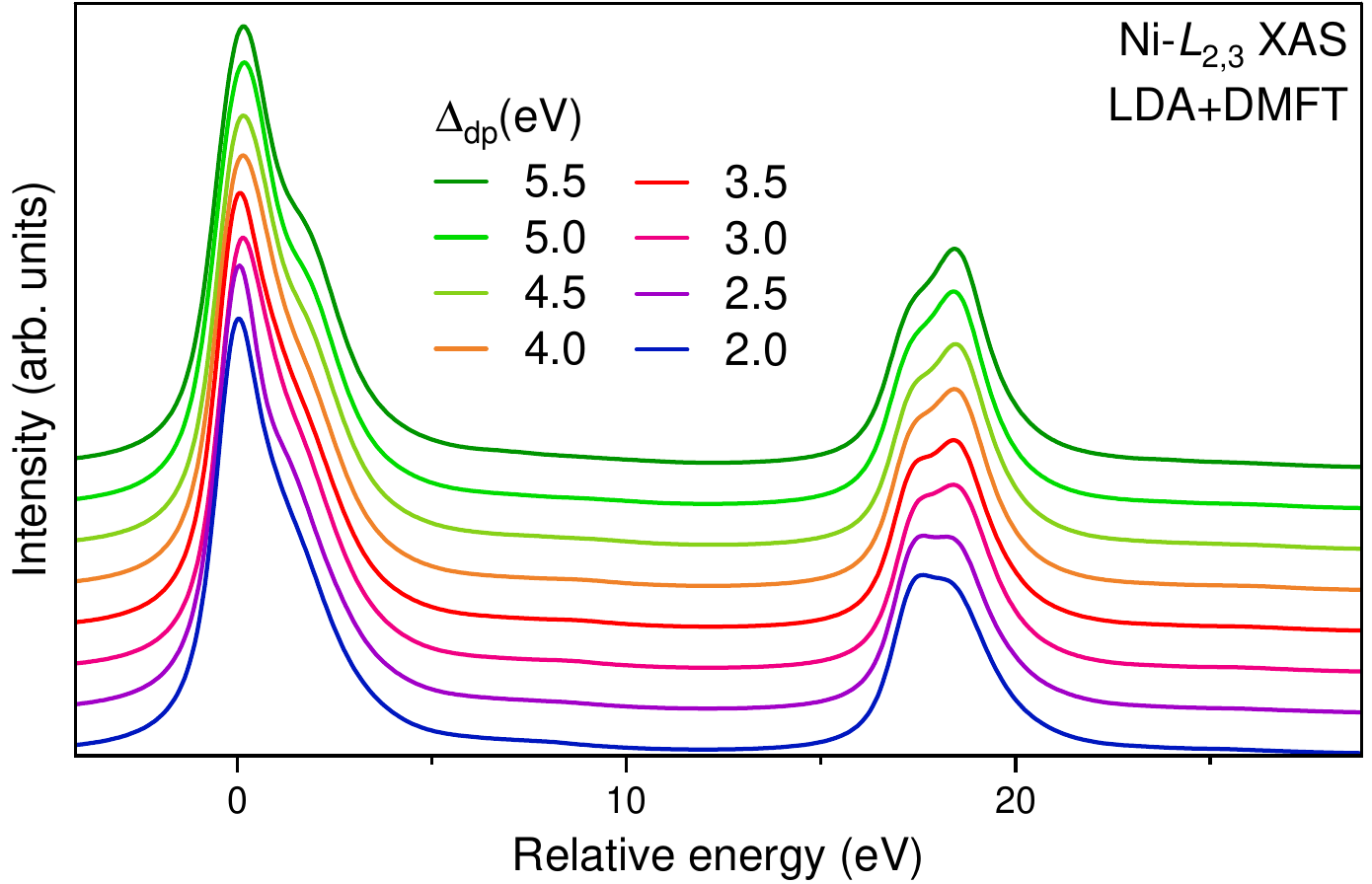}
\end{center}
\caption{LDA+DMFT theoretical Ni-$L_{2,3}$ X-ray absorption spectra of Nd$_3$Ni$_2$O$_7$.}
\label{Fig_XAS}
\end{figure}

\begin{figure}[]
\begin{center}
\includegraphics[width=0.55\columnwidth]{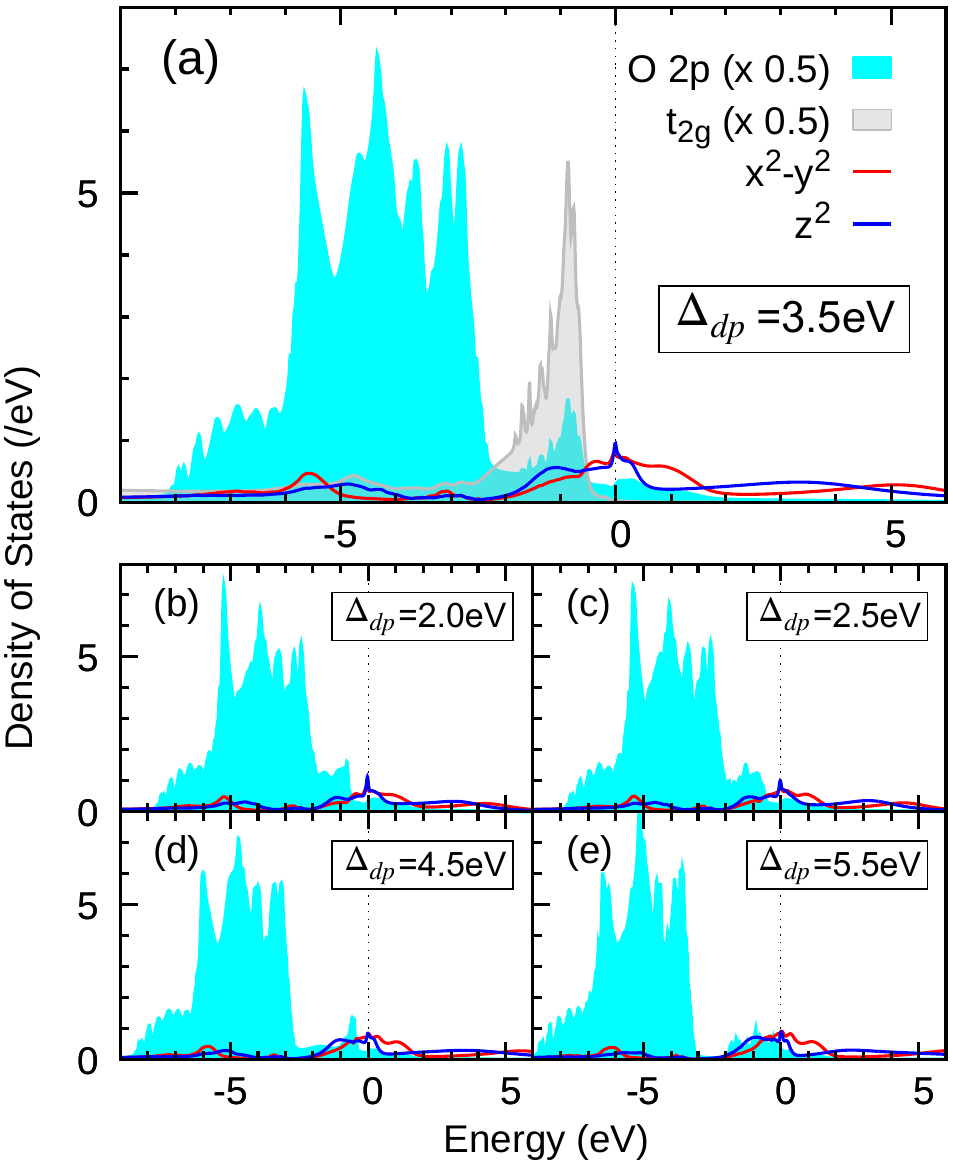}
\end{center}
\caption{The LDA+DMFT valence-band spectral intensities of Nd$_3$Ni$_2$O$_7$ calculated for selected $\Delta_{dp}$ parameters. Note that the spectral intensities of the O 2$p$ and Ni $t_{2g}$ states are scaled by a factor of 0.5 in the  panels.}
\label{Fig_VB_dmft}
\end{figure}

\begin{figure}[]
\begin{center}
\includegraphics[width=0.73\columnwidth]{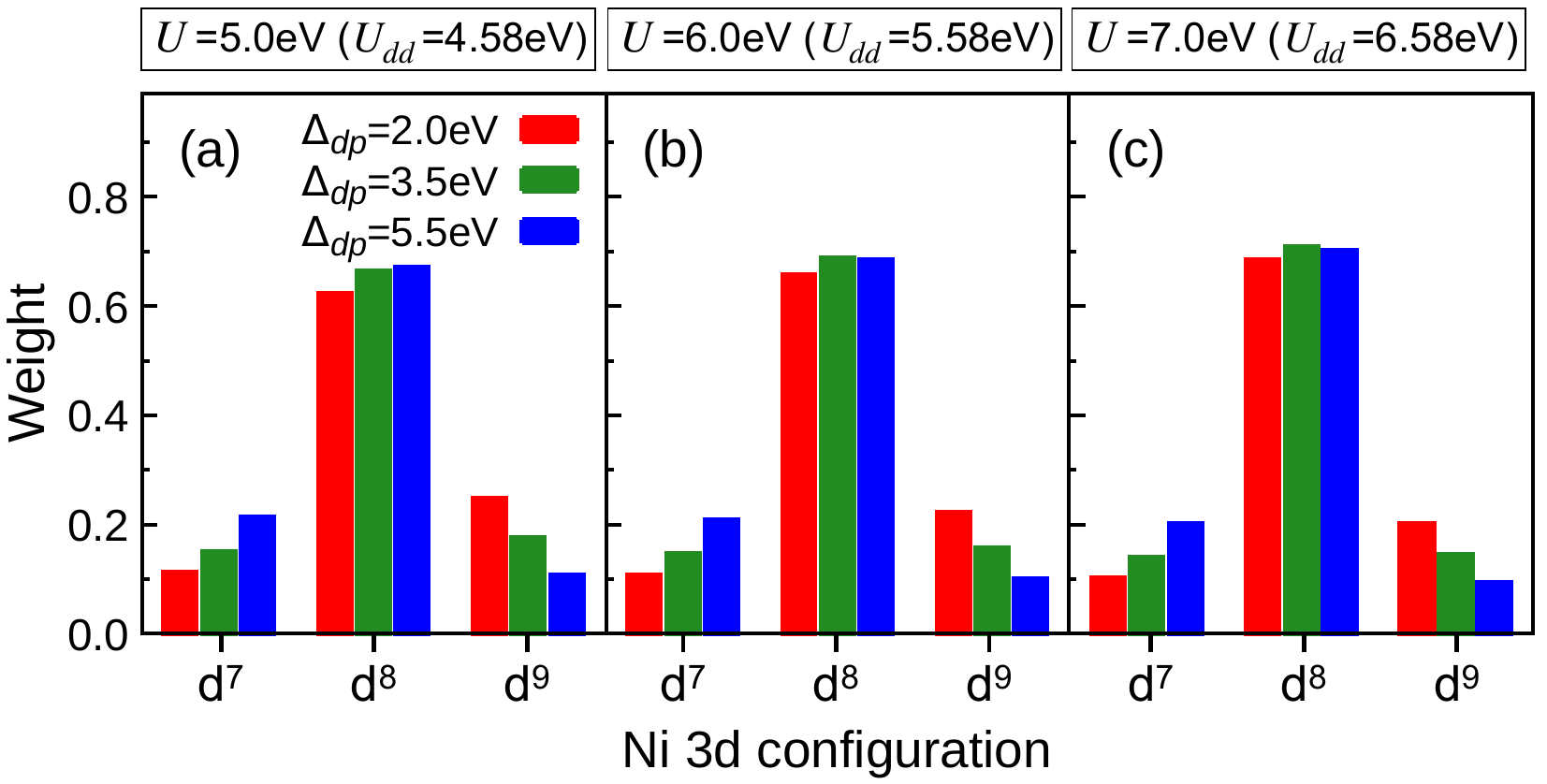}
\end{center}
\caption{The atomic histograms of the atomic configurations in Nd$_3$Ni$_2$O$_7$ calculated for the selected $\Delta_{dp}$ parameters with (a) $U=5.0$~eV, (b) $U=6.0$~eV, and (c) $U=7.0$~eV in the LDA+DMFT calculations.}
\label{Fig_hist_dmft}
\end{figure}

\begin{figure}[]
\begin{center}
\includegraphics[width=0.73\columnwidth]{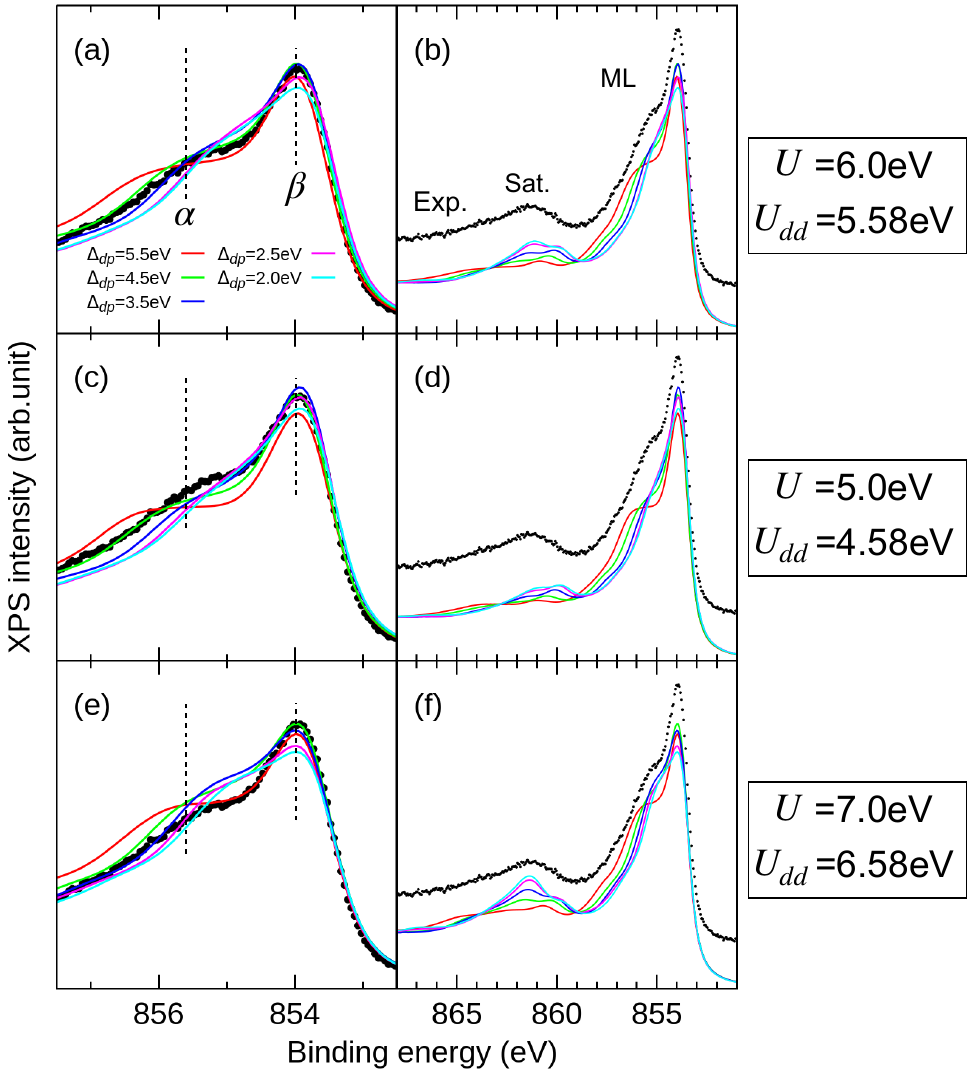}
\end{center}
\caption{(left) The Ni 2$p_{3/2}$ main line (ML) and (right) the full Ni 2$p_{3/2}$ spectrum including the satellite (Sat.), calculated by the LDA+DMFT AIM method using the different Hubbard $U$ parameters for a range of $\Delta_{dp}$ values. The experimental HAXPES spectrum is also shown for comparison.}
\label{Fig_XPS_dmft}
\end{figure}

Figure~\ref{Fig_XAS} shows the Ni \textit{L}$_{2,3}$ X-ray absorption spectra (XAS) of Nd$_3$Ni$_2$O$_7$ calculated for selected $\Delta_{dp}$ parameters. Both the \textit{L}$_{3}$ and \textit{L}$_{2}$ edges display two distinct features, with the ratio and their energy separation affected by $\Delta_{dp}$. The qualitative trends observed are similar to those experimentally observed in the Nd$_{n+1}$Ni$_n$O$_{3n+1}$ series~\cite{Pan2022}, with a reduction (increase) of the $\Delta_{dp}$ value resulting in a spectra closer to the divalent (trivalent) cases. 


\begin{figure}[]
\begin{center}
\includegraphics[width=0.85\columnwidth]{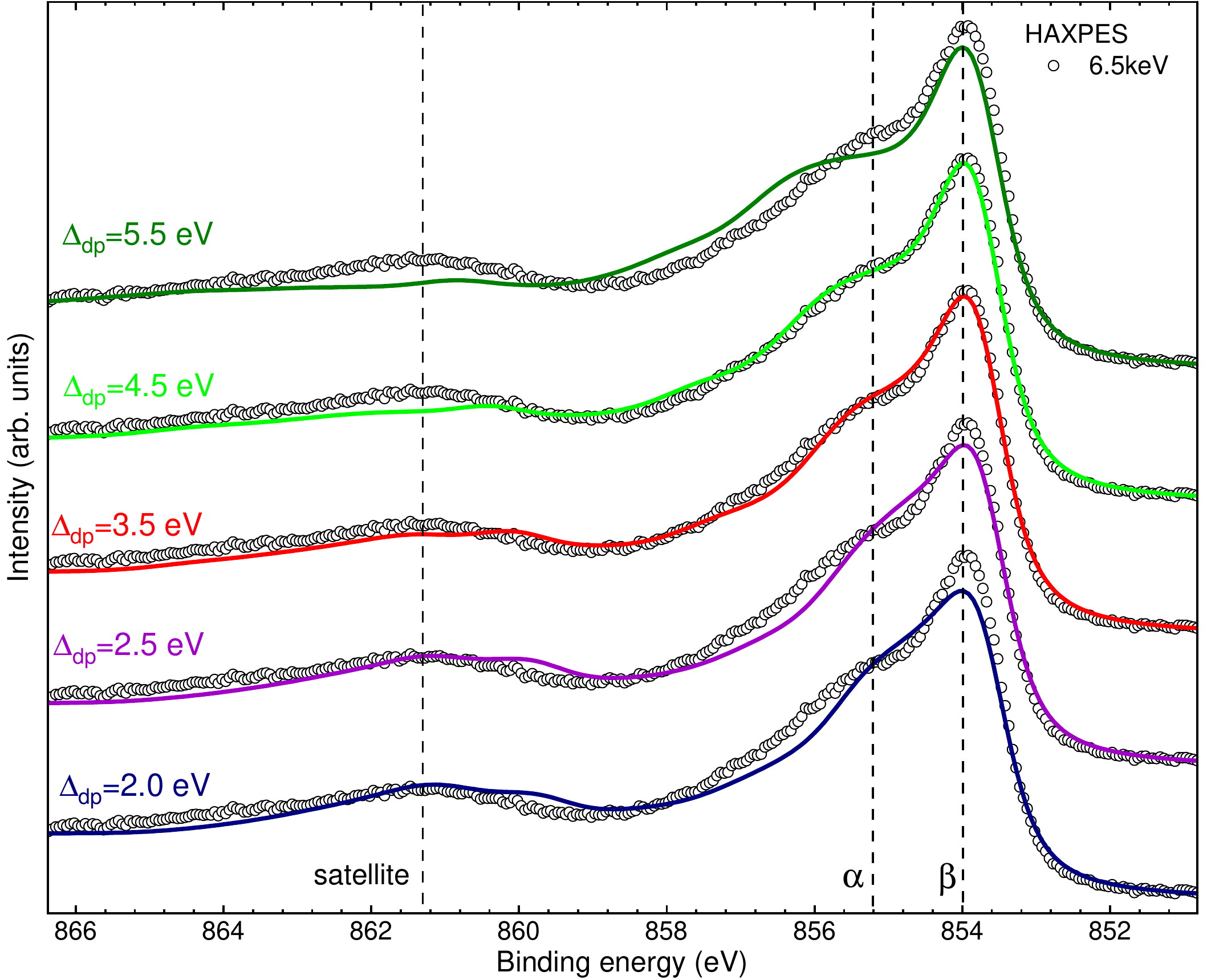}
\end{center}
\caption{Enlarged, individual comparison of the Ni 2$p_{3/2}$ LDA+DMFT calculations together and experimental HAXPES spectrum shown in Fig.~2(a,b).}
\label{Fig_2p_large}
\end{figure}
